\documentclass[11pt,a4paper,notoc]{article}
\pdfoutput=1
\usepackage{jcappub}

\usepackage[normalem]{ulem}

\DeclareMathOperator\erf{erf}

\begin{document}

\title{A new halo-independent approach to dark matter direct detection analysis}

\arxivnumber{OUTP-14-02p}

\author{Brian Feldstein}
\author{and Felix Kahlhoefer}
\affiliation{Rudolf Peierls Centre for Theoretical Physics, University of Oxford, 1 Keble Road, Oxford OX1 3NP, United Kingdom}

\emailAdd{brian.feldstein@physics.ox.ac.uk}
\emailAdd{felix.kahlhoefer@physics.ox.ac.uk}

\date{\today}

\abstract{
Uncertainty in the local dark matter velocity distribution is a key difficulty in the analysis of data
from direct detection experiments.  Here we propose a new approach for dealing with this
uncertainty, which does not involve any assumptions about the structure of the dark matter halo.
Given a dark matter model, our method yields the velocity distribution which best describes a set of direct detection data as a finite sum of streams with optimised speeds and densities.
The method is conceptually simple and numerically very efficient.  We give an explicit example
in which the method is applied to determining the ratio of proton to neutron couplings of dark matter from a hypothetical set of future data.
}

\keywords{Dark matter detectors, Dark matter experiments, Dark matter theory, Galaxy dynamics}

\notoc

\maketitle

\section{Introduction}

The direct search for dark matter (DM) in shielded underground detectors is a promising strategy not only for confirming the particle nature of DM, but also for measuring its key properties, such as mass and couplings to nucleons. A central problem in the analysis of such experiments, however, is the uncertainty in the DM velocity distribution $f(\mathbf{v})$~\cite{Kamionkowski:1997xg,Green:2010gw, McCabe:2010zh}. Numerical simulations indicate that assuming an isotropic and isothermal halo may not be a good approximation~\cite{Kuhlen:2009vh} and that in addition there may be localised streams of DM~\cite{Diemand:2008in} as well as a  DM disk co-rotating with the stars~\cite{Read:2009iv}. 

Such uncertainties are particularly important for light DM, which only probes the tail of $f(\mathbf{v})$~\cite{Lisanti:2010qx, Fairbairn:2012zs}, but they also significantly reduce the amount of information that can be inferred about general DM candidates. In fact, since changes in the halo structure can mimic changes in the DM parameters, a single direct detection experiment is insufficient to determine even the DM mass \cite{Drees:2008bv}. To make progress, one has to find a way to combine information from different target materials and quantify the impact of astrophysical uncertainties~\cite{Pato:2010zk, Pato:2011de, Peter:2013aha}.

The standard approach to this problem is to parametrize the uncertainties in $f(\mathbf{v})$ and scan (or marginalise) over the associated parameters~\cite{Strigari:2009zb, Peter:2009ak, Peter:2011eu, Arina:2011si, Pato:2012fw, Kavanagh:2012nr, Kavanagh:2013wba} (for alternative strategies see~\cite{Drees:2007hr, Drees:2008bv, Fox:2010bz}). This approach suffers from the problem that it is unclear whether the chosen parameterisation of the halo is sufficiently general. Moreover, direct detection experiments do not probe $f(\mathbf{v})$ directly, but instead probe the velocity integral $g(v_\text{min}) = \int_{v>v_\text{min}}   \hspace{-1mm} f(\mathbf{v}) / v \, \mathrm{d}^3\mathbf{v} $. Therefore, it is typically necessary to perform large numbers of numerical integrations over the DM velocity distribution, making this method numerically slow.  
Efficient scans then often require a Bayesian approach, with the need to motivate prior distributions for DM parameters.

In this letter we propose a new method for dealing with uncertainties in the DM velocity distribution. Instead of parametrizing $f(\mathbf{v})$, we directly parametrize $g(v_\text{min})$, so that predicted event rates depend on the parameters in a very simple way. In analogy to the treatment of the DM speed distribution $f(v) = \int f(\mathbf{v}) \mathrm{d}\Omega_v$ in~\cite{Peter:2009ak, Peter:2011eu}, we will write the velocity integral as a sum of step functions. This approach allows us to have a very large number of free parameters and removes the need to make any assumptions about the form of $f(\mathbf{v})$. Consequently, any conclusions drawn from it will be completely robust in the face of astrophysical uncertainties.  Moreover, our method involves a frequentist rather than a Bayesian approach, so that no prior distributions for any of the parameters need to be proposed.

\section{General framework for direct detection}
 
At a given experiment, the differential event rate with respect to recoil energy $E_\text{R}$ is given by:
\begin{equation}
\frac{\text{d}R}{\text{d}E_\text{R}} =  
\frac{C_\text{T} (A,Z) \, F^2(E_{\text{R}})}{2 \, \mu_{n\chi}^2} \, \tilde{g}(v_\text{min}) \; ,
\label{eq:dRdE}
\end{equation}
where $F(E_\text{R})$ is the appropriate nuclear form factor (taken from~\cite{Lewin:1995rx}) and $\mu_{n\chi}$ is the reduced DM-nucleon mass.  $C_\text{T}(A, Z) \equiv \left[Z \, f_p / f_n  + (A-Z)\right]^2$ with $A$ and $Z$ being the mass and charge number of the target nucleus, $f_p / f_n$ denoting the ratio of DM-proton to DM-neutron couplings and we have assumed spin independent scattering for simplicity.  

The \emph{minimum} velocity required for an energy transfer of size $E_\text{R}$ is given by $v_\text{min}(E_\text{R}) = \sqrt{m_\text{N} E_\text{R} / (2 \mu^2)}$, where $m_\text{N}$ is the mass of the target nucleus and $\mu$ is the reduced mass of the DM-nucleus system. The \emph{rescaled velocity integral}~\cite{Fox:2010bz, Frandsen:2011gi} is given by:
\begin{equation}
\tilde{g}(v_\text{min}) = \frac{\rho \, \sigma_n}{m_\chi} \, \int_{v > v_\text{min}} \frac{f(\mathbf{v})}{v} \,\text{d}^3\mathbf{v} \; ,
\label{eq:gtilde}
\end{equation}
where $\sigma_n$ is the DM-neutron scattering cross section, $m_\chi$ is the DM mass, $f(\mathbf{v})$ is the local DM velocity distribution in the lab frame and we take $\rho = 0.4~\text{GeV\,cm}^{-3}$ for the local DM density. In the Standard Halo Model (SHM), $f(\mathbf{v})$ in the Galactic rest frame is taken to be a truncated Maxwell-Boltzmann distribution with velocity dispersion $\sigma_\text{dis}= \sqrt{3/2} \times 220$~km\,s$^{-1}$ and escape velocity $v_{\rm{esc}}=544$~km\,s$^{-1}$.

For a detector with energy resolution $\Delta E_\text{R}$, the predicted number of events in a bin with energy range $\left[E_i, E_{i+1}\right]$ is given by~\footnote{A background contribution may be added as needed.}:
\begin{equation}
P_i = \int \kappa_i (E_\text{R}) \, \epsilon_\text{eff}(E_\text{R}) \, \frac{\text{d}R}{\text{d}E_\text{R}} \, \text{d}E_\text{R} \;,
\label{eq:Pi}
\end{equation}
where $\epsilon_\text{eff}(E_\text{R})$ is the effective exposure after including detector acceptance and
\begin{equation}
\kappa_i(E_\text{R}) = \frac{1}{2}\left[\erf\left(\frac{E_{i+1} - E_\text{R}}{\sqrt{2} \Delta E_\text{R}}\right)-\erf\left(\frac{E_i - E_\text{R}}{\sqrt{2} \Delta E_\text{R}}\right)\right]
\end{equation}
is the detector response function~\cite{Savage:2008er}.

\section{Finding the optimum velocity integral}
 
We observe from Eq.~\ref{eq:dRdE} that differential event rates depend on $f(\mathbf{v})$ in the same way for all experiments, namely through the monotonically decreasing function $\tilde{g}(v_\text{min})$~\footnote{Strictly speaking, this is true only if DM interactions with nuclei are velocity independent, as is the case in most models. Considering velocity dependent interactions is beyond the scope of this work (see~\cite{DelNobile:2013cva} for a discussion).}. When discussing astrophysical uncertainties it is therefore more efficient to work directly with the velocity integral~\cite{Fox:2010bz, Frandsen:2011gi, Gondolo:2012rs}. 

\begin{figure}[t]
\begin{center}
\includegraphics[width=0.55\columnwidth]{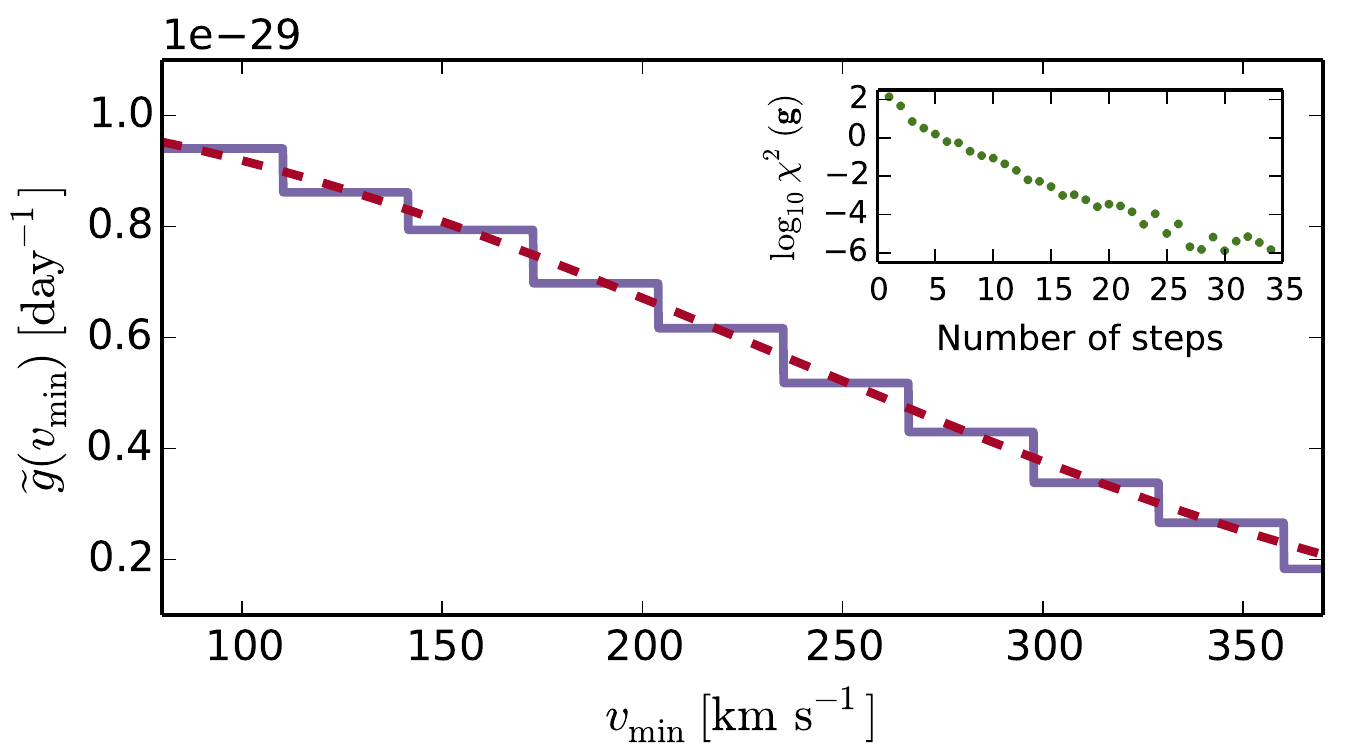}
\end{center}
\caption{
The red (dashed) line shows the SHM prediction for $\tilde{g}(v_\text{min})$, the purple line shows our best-fit approximation using 10 steps. Increasing the number of steps leads to a rapid decrease of $\chi^2(\mathbf{g})$, which quickly converges to zero (see inset).}
\label{fig1}
\end{figure}

Our goal is to find the function $\tilde{g}(v_{\rm min})$ that best describes a given set of data from direct detection experiments for a proposed DM model. We want to make \emph{no assumptions on the functional form} of $\tilde{g}(v_\text{min})$ apart from it being monotonically decreasing~\cite{Fox:2010bu}. We therefore proceed as follows:
\begin{itemize}
\item{We determine the region of $v_\text{min}$-space probed by the experiments under consideration by converting recoil energies into $v_\text{min}$-values. We then divide this region into $N_\mathrm{s}$ evenly spaced intervals of the form $\left[v_j, \, v_{j+1}\right]$.}
\item{We take the ansatz that $\tilde{g}(v_\text{min})$ is a monotonically decreasing sum of $N_{\rm s}$ steps, i.e.~we introduce $N_\mathrm{s}$ constants $g_j$, such that $\tilde{g}(v_\text{min}) = g_j$ for $v_\text{min} \in \left[v_j, \, v_{j+1}\right]$ (see Fig.~\ref{fig1}). The monotonicity requirement implies that $0 \leq g_j \leq g_{j-1}$ for all $j$~\footnote{Note that we take the overall normalisation of $\tilde{g}(v_\text{min})$ to be unconstrained, because in any case we allow $\sigma_n$ to take arbitrary values for a given $m_\chi$. For more restricted model hypotheses, we would have to include an additional constraint on the normalisation of $\tilde{g}(v_\text{min})$.}.}
\item{With this ansatz, and defining $E_j$ implicitly by $v_j = v_\text{min}(E_j)$, Eq.~\ref{eq:Pi} can be written in the simple form $P_i(\mathbf{g}) = \sum_j C_{ij} \, g_j$, where
\begin{equation}
C_{ij} = \frac{C_\text{T} (A,Z)}{2 \, \mu_{n\chi}^2} \int_{E_j}^{E_{j+1}} \hspace{-1.5mm} \kappa_i (E_\mathrm{R}) \, \epsilon_\text{eff}(E_\mathrm{R}) \, F^2(E_\mathrm{R}) \, \text{d}E_\mathrm{R}
\end{equation}
are calculable constants that depend only on the experimental details and the assumed DM properties, but are independent of astrophysics.}
\item{We now define the usual $\chi^2$ test statistic
\begin{equation}
\chi^2(\mathbf{g}) = \sum_i \frac{(P_i(\mathbf{g}) - N_i)^2}{P_i(\mathbf{g})}
\end{equation}
where $N_i$ is the experimentally observed number of events in the $i$th bin.}
\item{Next, we numerically find the step heights $g_j$ that minimise $\chi^2(\mathbf{g})$, maintaining the requirement of monotonicity $0 \leq g_j \leq g_{j-1}$. Although $N_\mathrm{s}$ can be rather large (see below), it is nevertheless easily possible to find the global minimum of $\chi^2(\mathbf{g})$. The reason is that the allowed region of $\mathbf{g}$ is convex and the Hessian of $\chi^2(\mathbf{g})$ is positive semi-definite everywhere within. Consequently, any local minimum of $\chi^2(\mathbf{g})$ is automatically a global minimum.}
\item{Finally, we take $N_\mathrm{s} \rightarrow \infty$. In practice, we take $N_\mathrm{s}$ so large that further increases yield negligible improvements to $\chi^2$. In typical examples of interest, we find that $N_\mathrm{s} \gtrsim 30$ is sufficient for this purpose. For such values of $N_\mathrm{s}$ the minimisation takes roughly a few seconds on a standard desktop computer.}
\end{itemize}
Effectively, we decompose the DM velocity distribution into a large number of streams with different densities and speeds. Since any continuous function may be approximated arbitrarily well by a sum of step functions, this method effectively finds the \emph{best possible form} for $\tilde{g}(v_{\rm min})$ for a given set of data and model parameters.

Interestingly, we have found that, even in the limit that $N_\mathrm{s} \rightarrow \infty$, the number of \emph{non-zero} steps in the optimised halos (i.e.~the number of steps with $g_j \neq g_{j+1}$) always remains smaller than the number of bins of data. This follows from the fact that, if we divide an optimised $\tilde{g}(v_{\rm min})$ into ``flat" sections with differing heights $h_i$, then these must all satisfy $\partial \chi^2 / \partial h_i =0$. 
It may be checked that these equations can only all be satisfied if either the number of flat sections is smaller than the number of bins, or if the predictions $P_i$ can be made to perfectly match the observations $N_i$. The latter possibility, however, is extremely unlikely for realistic cases including Poisson fluctuations. As a result, taking the limit $N_\mathrm{s} \rightarrow \infty$ does not actually lead to more and more steps being added to the optimised halos, but rather allows for finer adjustments of the \emph{endpoints} of the flat sections. Taking large $N_\mathrm{s}$ thus turns out to be a useful trick for determining the optimal endpoints for the flat sections in a numerically efficient way. 

We can now repeat our procedure for different sets of DM parameters and thereby find the minimum of $\chi^2(\mathbf{g})$, called $\hat{\chi}^2$, for every point in parameter space. The best-fit values for the DM parameters are then determined by finding the global minimum of $\hat{\chi}^2$. We can then define $\Delta \chi^2 = \hat{\chi}^2 - \hat{\chi}^2_\text{min}$, which we have confirmed to follow a $\chi^2$-distribution with the degrees of freedom given by the number of fitted DM parameters. Consequently, $\Delta \chi^2$ can be used to define confidence intervals as usual.

\begin{figure}[t]
\begin{center}
\includegraphics[width=0.49\columnwidth]{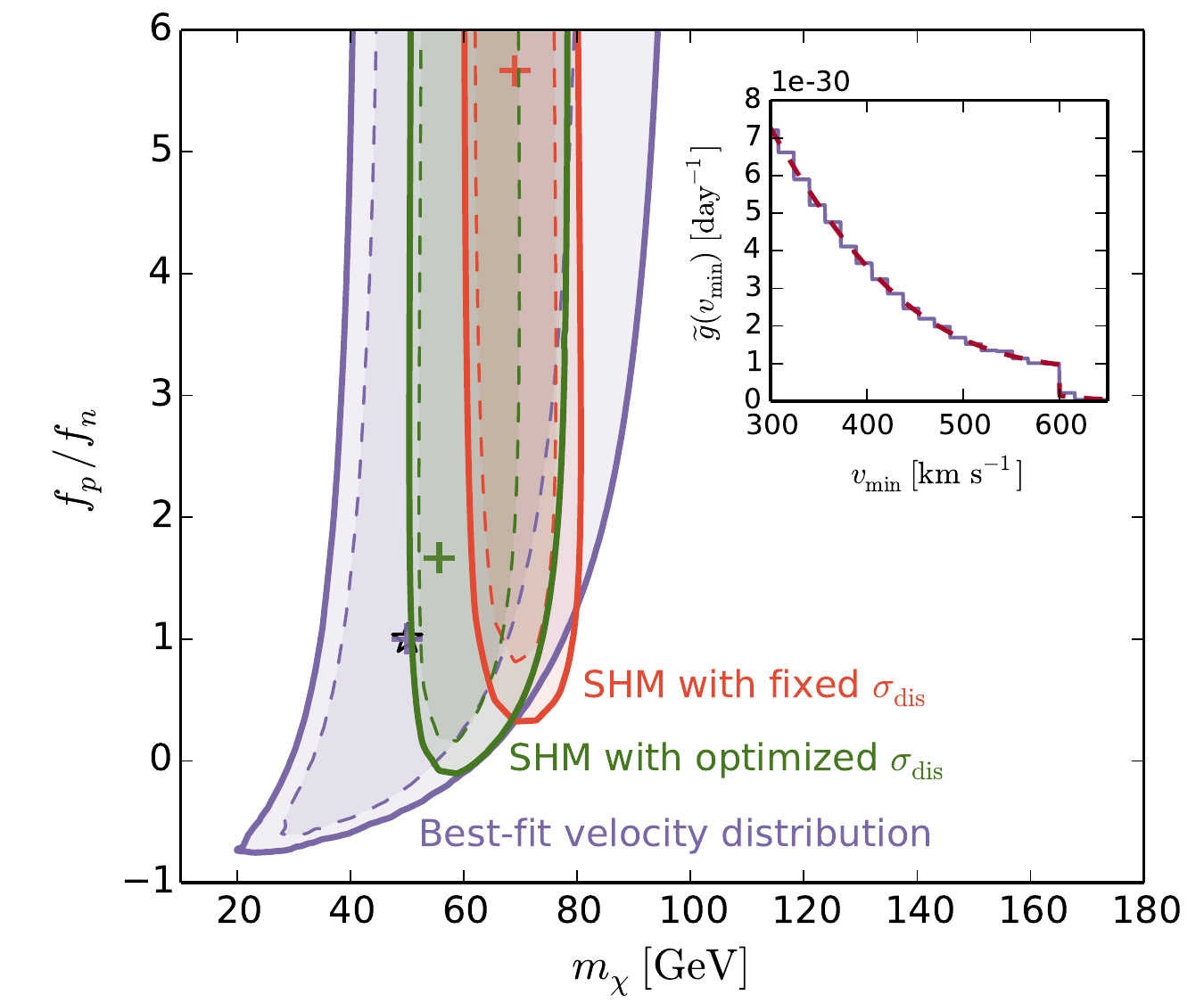}\hfill
\includegraphics[width=0.49\columnwidth]{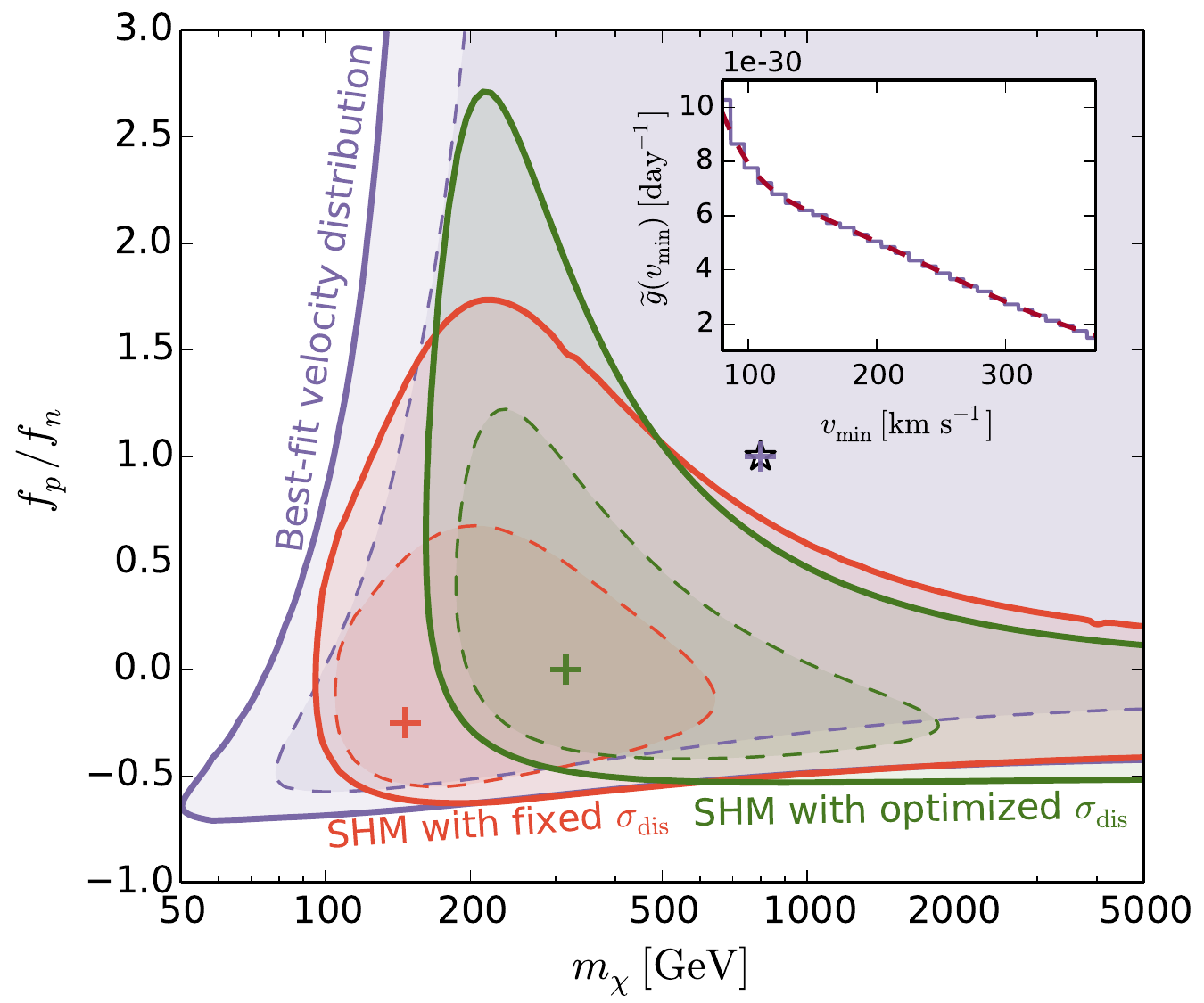}
\end{center}
\caption{
DM parameter estimation using mock data for different non-standard DM velocity distributions. For the left plot it is assumed that $10\%$ of the local DM density is in a DM stream, for the right plot it is assumed that $25\%$ of the local DM density is in a dark disk. The shaded regions correspond to the 68\% confidence regions (dashed contours) and 90\% confidence regions (solid contours) for three different methods (see text). The crosses indicate the corresponding best-fit values. The best-fit values for the DM parameters obtained from our new method (purple cross) perfectly coincide with the true values (white star). The insets show the best-fit halo that we obtain at the global minimum of $\chi^2(\mathbf{g})$ (solid purple line), compared to the assumed halo used to generate the data (red dashed line).}
\label{fig2}
\end{figure}

To illustrate this method, we have generated mock data for a set of future experiments (taken from~\cite{Pato:2010zk}, see below). Two particular examples are shown in Fig.~\ref{fig2}, where we use non-standard DM velocity distributions to generate a total of about 700 events across three different targets. For the left plot, we assume a $10\%$ contribution from a DM stream with a velocity of 600 km/s in the laboratory frame and take $m_\chi = 50~\text{GeV}$, $f_p / f_n = 1$ and $\sigma_n = 8 \times 10^{-46}~\text{cm}^2$. For the right plot we consider the case that the DM velocity distribution contains a $25\%$ contribution from a dark disk with velocity dispersion and lag relative to the baryonic disk both equal to $50~\text{km\,s}^{-1}$~\cite{Read:2009iv}, and have assumed $m_\chi = 800~\text{GeV}$, $f_p / f_n = 1$ and $\sigma_n = 7 \times 10^{-45}~\text{cm}^2$. In both cases we neglect Poisson fluctuations.\footnote{Note that the allowed parameter regions in both plots are unbounded, i.e.\ they extend to $f_p / f_n \rightarrow \pm \infty$. This is an artefact of the convention to show the coupling ratio rather than $\tan^{-1} f_p / f_n$.}

We observe that in both scenarios (as well as in all other cases that we have considered) the best-fit values for the DM parameters obtained from our method (purple cross) perfectly coincide with the true values (white star) and the best-fit $\tilde{g}(v_\text{min})$ matches perfectly the assumed velocity integral (see inset). For comparison, we also show the best-fit values and confidence regions obtained from two alternative methods, namely simply assuming the SHM with fixed parameters and optimising the fit over $\sigma_n$ only (orange) or assuming the SHM, but optimising the fits over both $\sigma_n$ and the velocity dispersion $\sigma_\text{dis}$ (green). It may be seen that both of these methods incorrectly exclude the true DM parameters at 90\% confidence level.

One observes from Fig.~\ref{fig2} that the contours obtained with our new method are rather broad, meaning that~-- although the assumed DM parameters are correctly reconstructed~-- the allowed parameter region is large. Clearly, tighter constraints on the DM parameters can be obtained by making stronger implicit or explicit assumptions about the DM halo while risking an incorrect exclusion of the true DM parameters for non-standard DM velocity distributions. The goal of our approach, however, is to determine DM parameters with no assumptions about this distribution, so that we only exclude DM parameters that cannot be made consistent with data by altering the DM halo. The success of our method is therefore based on yielding broader contours than other methods, not tighter ones.\footnote{It should be noted that other methods exist in the literature, such as the very general parameterisation of $f(v)$ discussed in Refs.~\cite{Kavanagh:2012nr, Kavanagh:2013wba, Kavanagh:2013eya}, that are well suited for reconstructing the DM parameters for non-standard velocity distributions. We leave a comparison of these methods with our approach to future work.}
 
\section{Example: Hypercharged Dark Matter}

We will now focus on a concrete model to show how our method is used in more detail.  If DM carries Standard Model hypercharge (and is in an appropriate representation of $SU(2)_L$ so that it has  an electrically neutral component), then it will generically interact with nuclei via tree-level $Z$-boson exchange, which results in a coupling ratio of $\sim -0.04$.  As discussed in~\cite{Feldstein:2013uha}, current direct detection constraints on such DM particles require masses greater than about $10^8~\text{GeV}$, and future experiments will be able to see a signal for $m_\chi$ up to $10^{10} - 10^{11}~\text{GeV}$. Such heavy DM particles most simply obtain an appropriate relic abundance by having masses close to the reheating temperature of the universe. Since the coupling strength of hypercharged DM is fixed, detection of a signal would immediately reveal the DM mass through the scattering rate, and this would then give otherwise unobtainable information about the thermal history of the universe.
In this scenario it would be of crucial importance to confirm that DM-nucleon scattering is mediated by $Z$-bosons. We will therefore use hypercharged DM as an example to show how our method may be used to determine $f_p / f_n$ in a halo-independent way.

The most likely target materials for ton-scale future direct detection experiments are Xenon, Germanium and Argon~\cite{Pato:2010zk}. Since their neutron to proton number ratios differ by less than 15\%, determining $f_p / f_n$ by looking at the relative rates of a signal on these elements will not be easy. Moreover, planned experiments using these elements will probe different regions of $v_\text{min}$-space, not only because the energy thresholds may differ, but also because heavier target nuclei are sensitive to smaller velocities.

Consequently, astrophysical uncertainties severely affect the determination of $f_p / f_n$.  If, for example, future data shows an excess of events in Xe-based experiments compared to Ar-based experiments, this observation could either be due to preferential coupling of DM to neutrons, as in the hypercharged scenario, or due to the DM velocity distribution decreasing more rapidly than in the SHM. Nevertheless, as long as there is non-negligible \emph{overlap} between the regions of $v_\text{min}$-space probed by different experiments, it will be possible to determine $f_p / f_n$ in a halo-independent way, given sufficient exposure.

\begin{figure}[t]
\begin{center}
\includegraphics[width=\columnwidth]{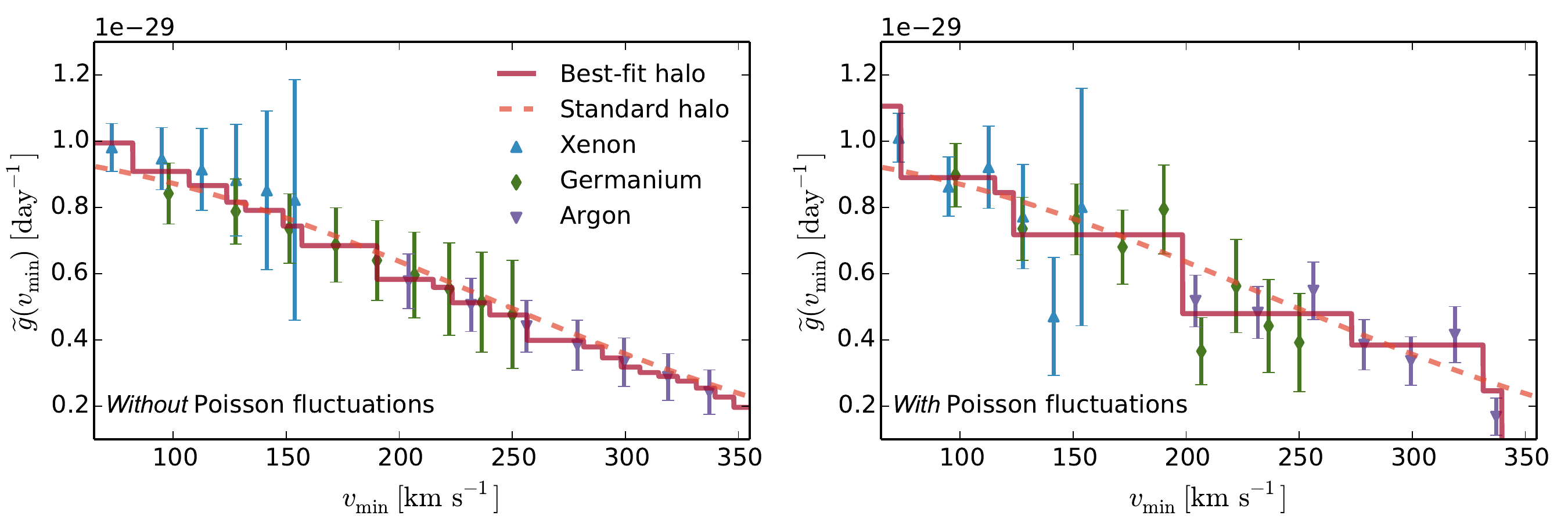}
\end{center}
\caption{Hypothetical measurements of the velocity integral $\tilde{g}(v_\text{min})$ from future experiments together with the best-fit velocity integral and the SHM velocity integral. For the top panel, the shown data points correspond to the prediction for hypercharged DM with $m_\chi = 7 \times 10^7~\text{GeV}$, but are interpreted under the \emph{incorrect} assumption that $f_p = f_n$. In the right panel, we show the same setup but additionally include a set of possible Poisson fluctuations.}
\label{fig3}
\end{figure}

This situation is illustrated in the top panel of Fig.~\ref{fig3}. The data points correspond to hypothetical measurements from future experiments employing Xe (blue), Ge (green) and Ar (purple) targets, assuming fermionic DM with hypercharge $1/2$ and $m_\chi = 7 \times 10^7~\text{GeV}$, compatible with the constraint from LUX~\cite{Akerib:2013tjd}, as well as a DM velocity distribution given by the SHM.  For this plot, we have chosen a bin width of $10$ keV and taken all experimental details from~\cite{Pato:2010zk}. In particular, we take energy-independent effective exposures of $2$, $2.16$, and $6.4$ ton-years and low energy thresholds of $10$, $10$ and $30$ keV at Xe, Ge and Ar, respectively.

For Xe we only include the lowest 6 bins, because its rapidly decreasing form factor cuts off the event rate at higher energies. We have checked that changing the number of bins or the bin size has negligible impact on our results. We map the experimental data onto $\tilde{g}(v_\text{min})$-values bin by bin using Eq.~\ref{eq:dRdE}~\cite{Fox:2010bz} and test the \emph{incorrect} model hypothesis that $f_p / f_n = 1$. 

Because of this incorrect hypothesis the data points from Xe predict values of $\tilde{g}(v_\text{min})$ that are large compared to the SHM prediction with best-fit normalisation (orange dashed line), while the Ar predictions are relatively small. The optimum velocity integral $\tilde{g}(v_{\rm min})$ as obtained from our method (red solid line), therefore clearly deviates from the SHM prediction and falls more steeply to reproduce this trend. In the absence of Poisson fluctuations, the value of $\hat{\chi}^2$ associated to the best-fit velocity integral in this example is $\hat{\chi}^2 = 1.05$.  Had we made the ``correct" hypothesis that $f_p / f_n = -0.04$, the best-fit velocity integral would be identical to the SHM prediction used to generate the data, giving $\hat{\chi}_\text{min}^2 \approx 0$ (see Fig.~\ref{fig1}).

The difference $\Delta \chi^2 = 1.05$ describes the extent to which $f_p / f_n = 1$ is disfavoured by the data. In this particular example, the hypothesis is excluded at the 69\% confidence level. Of course, Poisson fluctuations are expected to modify our conclusions. Nevertheless, our method still enables us to find the best-fit velocity integral even when the data points would favour an increasing velocity integral in some regions of $v_\text{min}$-space.  One possible example for the effect of such fluctuations is shown in the right panel of Fig.~\ref{fig3}, together with the corresponding best-fit velocity integral. If we include Poisson fluctuations, we find a median exclusion of the hypothesis that $f_p / f_n = 1$ at the $66\%$ confidence level. With a probability of about 22\%, the fluctuations are such that we can exclude $f_p / f_n = 1$ with at least 90\% confidence.

\begin{figure}
\begin{center}
\includegraphics[width=0.55\columnwidth]{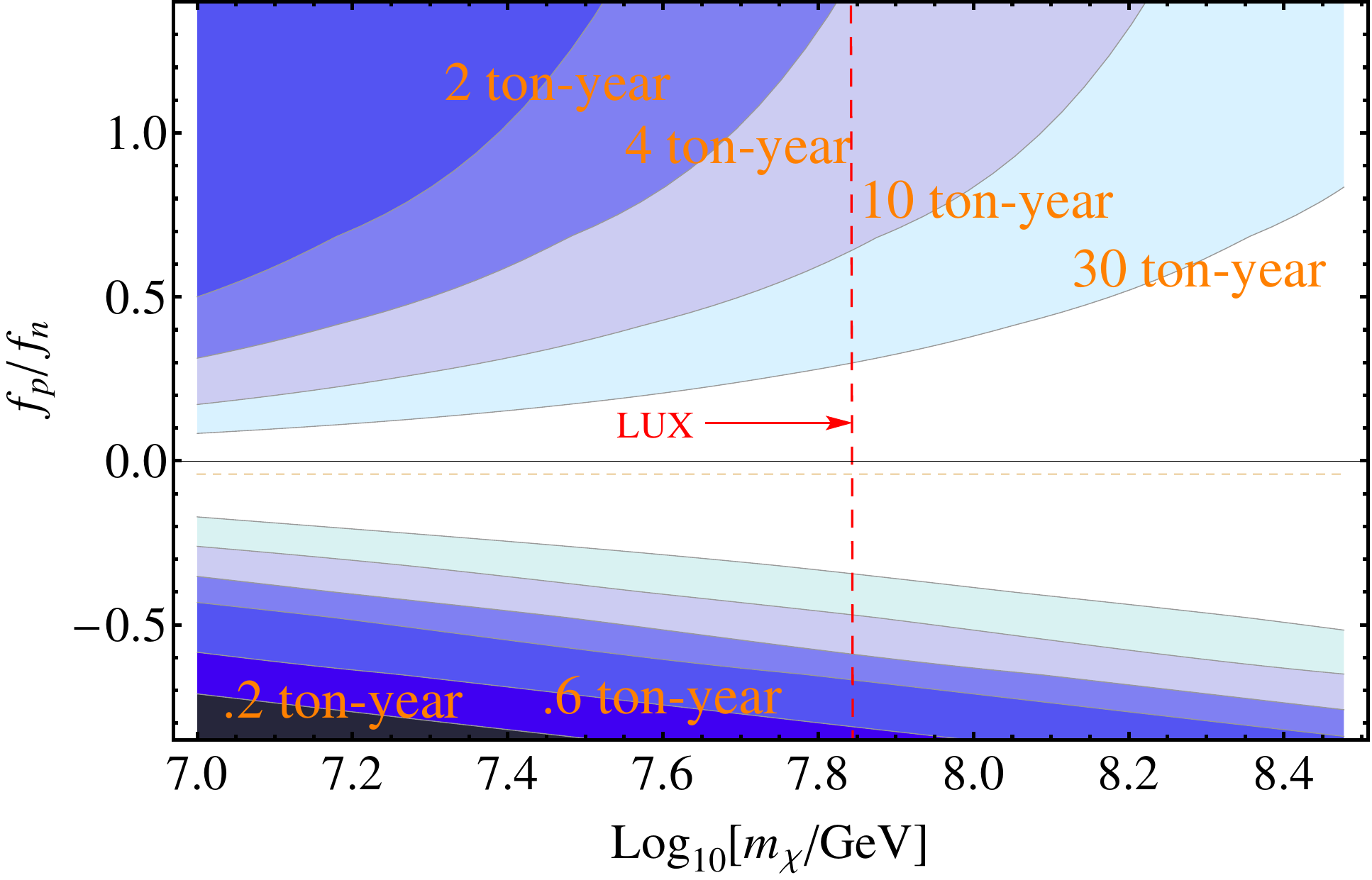}
\end{center}
\caption{Expected 90\% confidence intervals on   the ratio of DM couplings for different experimental exposures without making any assumptions about the DM velocity distribution. The underlying data was generated assuming hypercharged DM with mass given on the horizontal axis. The vertical dashed line indicates the lower bound on $m_\chi$ from the LUX experiment~\cite{Akerib:2013tjd}. 
}
\label{fig4}
\end{figure}

In Fig.~\ref{fig4} we show the $90\%$ confidence limits on $f_p / f_n$ which may be obtained from various exposures at Xe (rescaling the exposures at Ge and Ar from~\cite{Pato:2010zk} correspondingly~\footnote{Note that in contrast to~\cite{Feldstein:2013uha}, we are able to include Ar in our analysis.}). The data is generated assuming the same parameters for hypercharged DM as above, and we have minimised $\chi^2$ over both $\mathbf{g}$ and the hypothesised DM mass. In order to show what may be accomplished in a typical case, we have again ignored Poisson fluctuations for this plot.

To conclude this section, we note that the power to exclude $f_p / f_n = 1$ comes largely from the fact that Ge-based experiments have overlap in $v_\text{min}$-space with both Xe and Ar targets. In the absence of a Ge target, we find (neglecting Poisson fluctuations) $\Delta \chi^2 < 0.001$. We come to the important conclusion that for heavy DM essentially no information can be inferred about $f_p / f_n$ in a halo-independent way when using only Xe and Ar targets.

\section{Applications and future directions} The method we have presented is extremely general and given  a data set can be used to determine the best-fit velocity distribution  for a wide variety of possible model hypotheses. There is no obstacle to performing analyses of many more complicated particle physics models, such as inelastic~\cite{TuckerSmith:2001hy} or exothermic DM~\cite{Graham:2010ca, Essig:2010ye} or DM with long-range~\cite{DelNobile:2012tx} or momentum-suppressed~\cite{Chang:2009yt, Feldstein:2009tr} interactions. Similarly, our method could be applied to comparing annually modulating signals of DM, such as observed by DAMA~\cite{Bernabei:2010mq} and CoGeNT~\cite{Aalseth:2014eft}, and possibly also to constraining the modulation fraction.

Another exciting prospect is to apply our method to experimental results that give contradictory information when interpreted in terms of the SHM (such as LUX, CoGeNT, SuperCDMS~\cite{Agnese:2014aze} and CDMS-Si~\cite{Agnese:2013rvf}) to understand if a different DM velocity distribution can bring these experiments into agreement. Here, however, there are two important complications.
The first issue is that, for experiments with unclear compatibility, one would be interested in determining the goodness of fit at the actual best-fit point in addition to the determination of confidence intervals in DM parameter space.  Typically, the value of the $\chi^2$ statistic at a best fit point follows a $\chi^2$ distribution with number of degrees of freedom given by the number of observations (i.e.~bins) minus the number of free parameters. 
Our requirement of monotonicity, however, makes the notion of the number of free parameters in our halo fits somewhat unclear.  In particular, we know that it generally remains impossible for us to fit data arbitrarily well even in the limit of an infinite number of $\tilde{g}(v_{\rm min})$ steps.
The second issue is that, for experiments observing small event rates, binning the data and using $\chi^2$-methods becomes unreliable.  In principle, there is no obvious obstacle to using a binned or unbinned extended maximum likelihood method~\cite{Barlow1990496} to determine the optimum $\tilde{g}(v_\text{min})$, but doing so would appear to make future progress on the goodness-of-fit question difficult.\footnote{Though we note that such methods might be useful for DM parameter determination in cases with bins having fewer events than what we have assumed in this paper.} We leave these problems to future work.

\section*{Acknowledgements}

We thank Patrick Fox, Masahiro Ibe, Christopher McCabe and Matthew McCullough for useful discussions. BF is supported by STFC UK and FK is supported by the Studienstiftung des Deutschen Volkes and a Leathersellers' Company Scholarship at St Catherine's College, Oxford.

\providecommand{\href}[2]{#2}\begingroup\raggedright\endgroup

\end{document}